\DocumentMetadata{}
\documentclass[authorversion, sigconf]{acmart} 

\usepackage{booktabs}
\usepackage{multirow}
\AtBeginDocument{%
  }

\copyrightyear{2024} 
\acmYear{2024} 
\setcopyright{rightsretained} 
\acmConference[Augmented Humans 2025]{}{}{}

\begin{document}

\title{MaskClip: Detachable Clip-on Piezoelectric Sensing of Mask Surface Vibrations for Real-time Noise-Robust Speech Input}

\author{Hirotaka Hiraki}
\affiliation{%
  \institution{The University of Tokyo}
  \streetaddress{7-3-1, Hongo}
    \state{Tokyo}
  \country{Japan}
  \postcode{43017-6221}
}
\affiliation{%
  \institution{National Institute of Advanced Industrial Science and Technology}
    \state{Chiba}
  \country{Japan}
  \postcode{43017-6221}
}
\email{hirotakahiraki@gmail.com}
  
\author{Jun Rekimoto}
\affiliation{%
  \institution{The University of Tokyo}
  \streetaddress{7-3-1, Hongo}
  \state{Tokyo}
  \country{Japan}
  \postcode{43017-6221}
}
\affiliation{%
  \institution{Sony CSL Kyoto}
  \streetaddress{7-3-1, Kyoto}
  \state{Kyoto}
  \country{Japan}
  \postcode{43017-6221}
}
\email{rekimoto@acm.org}

\renewcommand{\shortauthors}{Trovato et al.}


\begin{teaserfigure}
  \includegraphics[width=\textwidth, page=1]{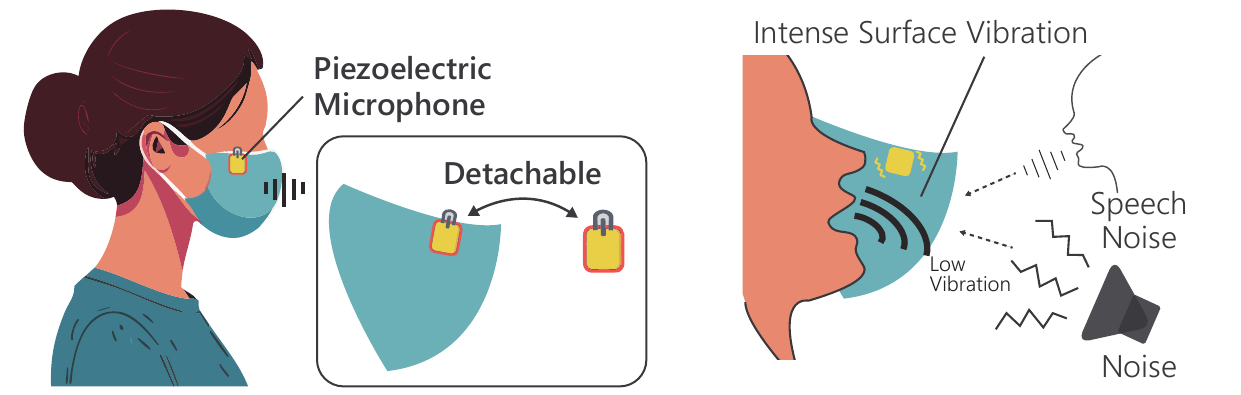}
  \caption{Overview of the MaskClip system architecture and operational principles. The left illustration shows the device in its deployed state, while the center diagram details the detachable clip-on design. The right illustration demonstrates the speech transmission mechanism, showing how mask surface vibrations are detected while ambient noise is attenuated. The piezoelectric microphone preferentially captures mask surface vibrations, effectively filtering out environmental sounds. This design enables superior speech recognition performance in noisy environments such as medical settings. The detachable nature of the device maintains hygiene standards while ensuring consistent voice capture quality.} 
  \label{fig:teaser}
\end{teaserfigure}


\begin{abstract}


Masks are essential in medical settings and during infectious outbreaks but significantly impair speech communication, especially in environments with background noise. Existing solutions often require substantial computational resources or compromise hygiene and comfort. We propose a novel sensing approach that captures only the wearer's voice by detecting mask surface vibrations using a piezoelectric sensor. Our developed device, MaskClip, employs a stainless steel clip with an optimally positioned piezoelectric sensor to selectively capture speech vibrations while inherently filtering out ambient noise. Evaluation experiments demonstrated superior performance with a low Character Error Rate of 6.1\% in noisy environments compared to conventional microphones. Subjective evaluations by 102 participants also showed high satisfaction scores. This approach shows promise for applications in settings where clear voice communication must be maintained while wearing protective equipment, such as medical facilities, cleanrooms, and industrial environments.

\end{abstract}





\maketitle

\section{Introduction}

Face mask usage has become commonplace across various settings, from healthcare environments to everyday office spaces. 
Particularly during COVID-19 and influenza outbreaks, widespread mask-wearing significantly impacted communication effectiveness.
Studies have shown that masks cause significant speech attenuation ~\cite{Al-Karawi2023-zm,Saeidi2015-dm}, affecting both speech intelligibility and recognition accuracy. This challenge becomes more pronounced in noisy environments such as hospitals and industrial settings, where multiple conversations and equipment sounds create complex acoustic interference.

In modern healthcare environments, there is a growing expectation for voice-controlled medical devices and automated record-keeping systems ~\cite{Zelinka2010-rn,Schulte2020-me,Zhang2023-yu,Latif2021-wt}. However, the presence of masks in these settings presents significant challenges for voice input quality and reliability. This issue is further compounded by the inherently noisy environment of medical facilities and other professional settings.

Previous approaches to address these challenges include software-based noise suppression ~\cite{Sun2021-dl} and multimodal speech enhancement ~\cite{earse}. However, these methods often require substantial computational resources, making them impractical for resource-constrained devices. Earphone-based source separation methods ~\cite{clearbuds,Veluri2024-lp} can inadvertently treat frontal sound as noise. Existing mask-type microphones ~\cite{whispermask_ahs} pose hygiene concerns for long-term use due to their direct skin contact and difficulty in removal. 


%
Additionally, wearable silent speech solutions ~\cite{ReHEarSSE, E-mask, jawsense, SilentMask} prove inadequate for clear speech communication. Silent speech solutions refer to systems that capture non-audible speech articulations (such as lip movements, tongue positions, or facial muscle activity) to recognize speech without the user producing audible sounds. While these systems can work for limited command sets, they typically lack the expressiveness and vocabulary range needed for natural conversation and general-purpose speech recognition.

%


To overcome these limitations, we propose a hardware-based approach: a detachable piezoelectric sensor that captures the user's voice through mask surface vibrations(Fig \ref{fig:teaser}). This method builds upon the concept of throat microphones, which have been used in noisy environments ~\cite{Wang2023-ll,Erzin2009-pd}. However, traditional throat microphones suffer from high-frequency attenuation and susceptibility to spike noise during movement ~\cite{fukumoto}. Additionally, throat microphones typically require tight contact with the skin, causing friction between the sensor and skin during head movements, which produces disruptive spike noise in the audio signal. Our detachable piezoelectric-based approach mitigates these issues by capturing vibrations directly from the mask surface.

Our detachable clip-on design addresses hygiene concerns by allowing the device to be easily attached and removed without compromising mask integrity. This approach enables masks to be disposed of or cleaned separately from the sensing device, making it suitable for environments with strict hygiene requirements.

For evaluation, we developed a comprehensive test environment that simulates real-world usage conditions. Using the LibriMix and WHAM! datasets ~\cite{Zhang2021-hs}, we incorporated both speech noise from two speakers and background noise from a single speaker. 
Our experimental results demonstrate that the proposed system achieves a Character Error Rate (CER) of 6.1\% in noisy environments, compared to conventional pin microphones which exhibited 19.7\% CER in the same conditions. Additionally, subjective evaluations using the MUSHRA methodology with 102 participants showed that MaskClip achieved significantly higher audio quality scores (73.0±9.8 points) than conventional microphones (34.8±2.9 points) for normal speech in noisy environments. These results demonstrate the system's superior performance in both objective and subjective measures.

The key contributions of this research are:

\begin{itemize}
\item A hardware-based noise suppression approach utilizing mask surface vibration detection
\item Implementation of a hygiene-conscious detachable clip-on device design
\item Quantitative evaluation of speech recognition performance in complex noise environments
\item Demonstration of practical voice interface solutions applicable across various settings, from healthcare facilities to general office environments
\end{itemize}

\begin{figure*}
    \centering
    \includegraphics[width=\linewidth, page=2]{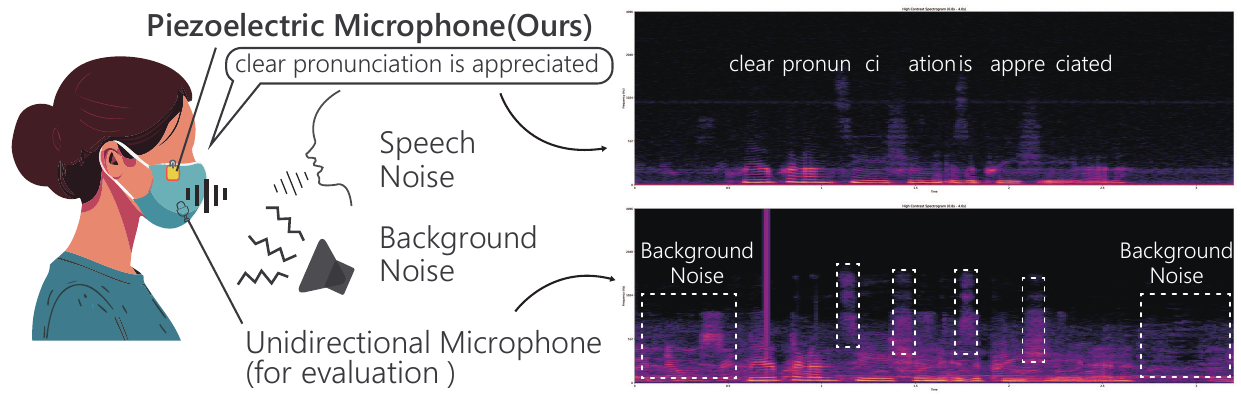}
    \caption{Comparison of speech recognition performance between MaskClip and a unidirectional microphone. The upper spectrogram shows the recording from MaskClip, demonstrating clear speech articulation. The lower spectrogram shows the recording from a conventional unidirectional microphone, revealing significant background noise interference. The contrast illustrates MaskClip's superior noise resistance capabilities.}
    \label{fig:abstract_maskclip}
\end{figure*}

\section{Related Work}

    
    

\subsection{Software Based Speech Enhancement and Speech Source Separation}

Clear audio recording is crucial for effective communication and preventing errors in voice input systems. Various software techniques have been developed to address this need. Methods such as Independent Component Analysis (ICA) ~\cite{ICA_bss}, Non-negative Matrix Factorization (NMF)~\cite{nmf_for_bss}, and Kalman filtering ~\cite{kalmanfilter} aim to extract clear audio by estimating noise characteristics. While these approaches are effective for specific frequency or stationary noises like power line interference or equipment noise, they struggle to separate overlapping human voices.
Recent advancements in deep learning have led to the development of techniques that extract audio features to enhance or separate speakers' voices. ConvTasNet~\cite{Conv-tasnet}, for example, transforms signals into the time-frequency domain and employs convolutional neural networks to separate noise from the target signal. This model is computationally efficient, allowing real-time inference on GPUs.
In addition to time-frequency domain approaches, methods that directly process raw audio in the time domain have been proposed. Denoiser~\cite{denoiser}, for instance, utilizes a Recurrent Neural Network (RNN) structure to enhance audio solely in the time domain, making it well-suited for streaming applications. However, RNN-based structures like Denoiser can be computationally intensive due to their reliance on past inference results for processing subsequent audio frames.
To address this limitation, Sepformer~\cite{sepformer} introduces a Transformer-based architecture for audio source separation. Its attention mechanism enables consistent inference over extended time sequences. Furthermore, in the field of deep learning-based speech recognition, recent developments in Large Language Models (LLMs) have led to techniques that reinterpret speech recognition results using LLMs~\cite{robustGER}. This approach aims to capture linguistic coherence and mitigate the effects of noise on recognition accuracy.

\subsection{Hardware Based Speech Enhancement}





In hardware-based solutions, various types of microphones are employed to reduce noise by combining their sound capture mechanisms with signal processing techniques. These include unidirectional microphones ~\cite{A-century-of-microphones}, throat microphones ~\cite{throatmic1, throatmic2}, Non-Audible Murmur (NAM) microphones ~\cite{nam_microphone}, and earphone-based microphones ~\cite{clearbuds}, each leveraging different principles of sound acquisition.

Unidirectional microphones, commonly found in lavalier and headset microphones, represent one of the most accessible noise reduction techniques today. By blocking the diaphragm’s rear side and restricting its vibration direction to one side, these microphones achieve a directional focus of 180 degrees ~\cite{A-century-of-microphones}. This allows for stronger recording of sounds from limited directions, making them suitable for scenarios like panel discussions where multiple speakers may talk simultaneously. However, since they do not limit the distance from which sound is captured, background noise cannot be eliminated by the microphone alone, making it challenging to use them in noisy environments.

Earphone-based microphones are designed to be inserted into the ears, with earpieces attached to both ears, enabling speaker selection through beamforming techniques. Machine learning methods have also been proposed to reduce noise by utilizing the audio captured by both ears ~\cite{clearbuds, In-Ear-Voice}. However, these microphones struggle with capturing less distinct voices, such as whispers.

Throat microphones use piezoelectric elements to convert the vibrations on the surface of the neck, generated during speech, into audio ~\cite{Throat-microphone}. By being attached to the neck and capturing only the vibrations of the skin surface, these devices can isolate the user’s voice while remaining highly resistant to background noise, as environmental sounds do not sufficiently vibrate the piezoelectric elements. However, since they need to be tightly attached to the neck, movements like head turns or nods can introduce noise. Additionally, because the captured voice passes through the body’s tissue, important speech features such as formants are lost ~\cite{DirectSpeechReconstruction}, necessitating post-processing to achieve intelligible and recognizable audio.

NAM microphones, similar to throat microphones, are worn on the skin behind the ear. However, they utilize omnidirectional microphones encased in silicone, smoothing the boundary between the microphone and the skin ~\cite{nam_microphone}. Like throat microphones, NAM microphones significantly reduce the influence of environmental noise. However, movements such as nodding can still introduce noise. Furthermore, while NAM microphones capture sound from behind the ear, post-processing such as speech enhancement remains necessary ~\cite{NAM0, NAM1, NAM2}.

\subsection{Multi Channel Speech Enhancement}
%
Non-waveform data associated with speech have also been highlighted as important approaches for enhancing speech signals. Methods using ultrasound, acoustic imaging, and radar have been proposed to capture fine movements of the lips during speech through the reflection of sound waves, thereby enhancing the speech signal.

Ozturk et al. introduced RadioSES ~\cite{RadioSES}, a method for sound source separation using millimeter-wave sensing. RadioSES captures the vibrations of a speaker’s vocal cords through a channel distinct from the voice signal, allowing for enhanced speech separation. However, this method faces challenges when applied to individuals in motion.

In systems such as UltraSE ~\cite{Sun2021-dl}, UltraSpeech ~\cite{UltraSpeech}, and EarSE ~\cite{earse}, ultrasound emitted from speakers embedded in smartphones or headphones is reflected off the user's face during speech, capturing the profile as supplementary information to the voice. This profile is then used to enhance the speech by reconstructing the non-noise profile. However, these systems require the user to speak directly towards the smartphone, making hands-free usage impractical. Additionally, there is significant latency in communication, as inference using the smartphone’s GPU takes more than 15 seconds for 5 seconds of audio input.

\begin{figure*}[htbp]
    \centering
    \includegraphics[width=\linewidth, page=3]{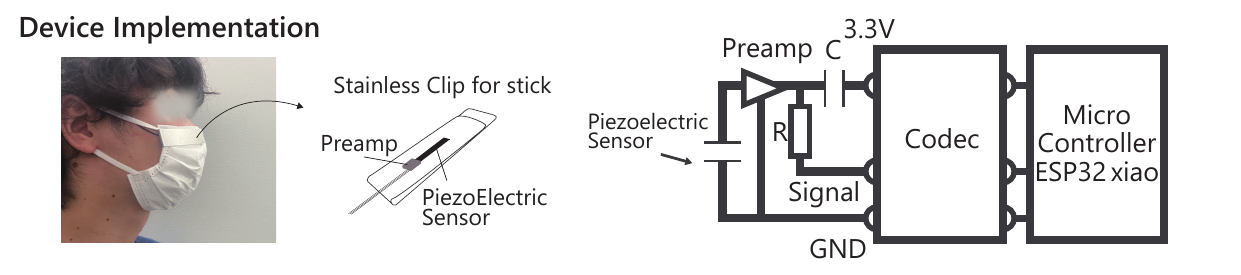}
    \caption{Detailed hardware implementation of the MaskClip device. The left image shows the physical device implementation, while the right diagram illustrates the electrical circuit configuration. The core components consist of a piezoelectric sensor, preamplifier circuit, codec, and ESP32 xiao microcontroller. The signal processing path operates at 3.3V, where signals from the piezoelectric sensor are amplified by the preamplifier, digitized by the codec, and processed by the microcontroller. The stainless-steel clip design ensures secure attachment and stable signal detection. This architecture enables low-power operation while maintaining high-performance voice detection capabilities. The compact form factor and integrated design demonstrate the system's practical viability for real-world applications.}
    \label{fig:maskclip_device}
\end{figure*}

\subsection{Mask-type Interface}
The increased use of masks during the COVID-19 pandemic has brought attention to the potential of masks as wearable interfaces. Notably, it has been reported that masks cause speech attenuation, making speech recognition more challenging ~\cite{mask-is-difficult-to-speech-recognition1, mask-is-difficult-to-speech-recognition2}. To address this issue, researchers have proposed methods that incorporate sensors into masks, enabling silent speech recognition ~\cite{SilentMask, E-mask}. However, these approaches primarily focus on recognizing specific commands for smart assistants such as Alexa, limiting their applicability to more comprehensive speech recognition or conversational capabilities.

Mask-integrated microphones have also been proposed, where flexible sensors are embedded within the mask~\cite{whispermask_chiEA,whispermask_ahs}. However, since each mask requires a dedicated sensor, these systems present challenges in terms of reuse while maintaining hygiene.

A significant issue with mask-wearing is the obstruction of facial expressions, which can lead to communication difficulties. Researchers have proposed methods to detect facial expressions using photoreflectors, capacitive touch sensors, and display them on LED screens ~\cite{unmasked, mascreen, TransEmotion, TexonMask}. Furthermore, masks have been adapted as wearable interfaces for various interactions related to the face and mouth, including breath detection ~\cite{ibuki, ReliableBreathing, Respiration}, gaze tracking ~\cite{SleepMask}, mouth shape recognition ~\cite{yadori, mouthgesture}, and mask donning/doffing detection using straps ~\cite{masktrap}.

\section{MaskClip}
MaskClip is designed based on the requirements for embedded interaction, considering masks as everyday wearable devices. Key requirements include hands-free operation, detachability, long-term usability, and maintaining the original comfort of the mask.
Hands-free operation allows simultaneous use with smartphones and other devices, adding voice interaction as an additional modality.
Detachability enables adaptation to different work scenarios and purposes. Simple attachment and detachment processes allow a wide range of users, from children to the elderly, to set up the device.
Maintaining the original comfort of the mask is crucial. Heavy components can cause the mask to slip, increasing the need for adjustments and potentially introducing noise. Based on these considerations, this study proposes MaskClip, a detachable piezoelectric mask microphone.

\subsection{Principle: sensing the surface of the mask}
MaskClip operates based on the piezoelectric effect to directly detect mask vibrations, implementing a detachable microphone through a clip equipped with a piezoelectric sensor. Conventional condenser microphones, such as those in headsets, contain an air-vibrating diaphragm inside the microphone that generates slight voltage changes through diaphragm vibration. These changes are amplified to capture voice input. In contrast, MaskClip uses a piezoelectric microphone where voltage is generated through deformation of the piezoelectric material when vibrations are applied. Voice input is achieved by sampling this voltage at audio sampling frequencies (e.g., 44.1 kHz).
Voice input becomes possible as vibrations transmit through the mask surface and clip. This mechanism predominantly captures sounds that induce mask movement while filtering out ambient noise, particularly noise that does not cause mask vibration. As demonstrated in Fig. \ref{fig:abstract_maskclip}, this approach achieves superior noise resistance compared to conventional unidirectional microphones, with spectrograms clearly showing the difference in speech clarity and background noise rejection. Using this principle, we propose a mask-type device that effectively captures the wearer's voice. While throat microphones employ a similar mechanism, they require constant sensor contact with the neck, resulting in spike noise during head movements~\cite{throatmic1, throatmic2}.

\begin{figure*}[htbp]
    \centering
    \includegraphics[width=\linewidth, page=4]{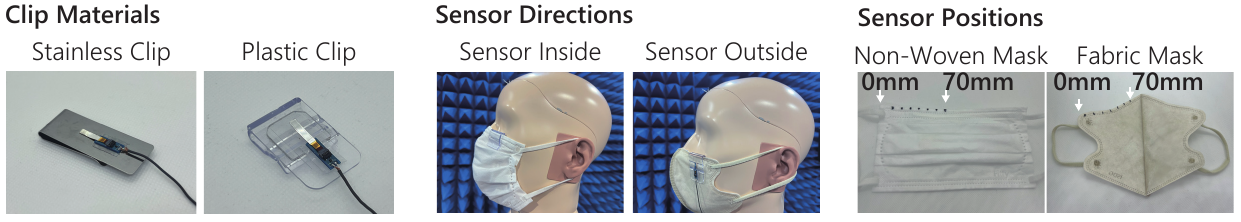}
    \caption{Overview of experimental conditions used for systematic evaluation of MaskClip's sensor configuration parameters. From left to right: clip materials (stainless steel versus plastic), sensor directions (inward-facing versus outward-facing), and positioning (non-woven mask versus fabric mask). Performance impacts were evaluated across combinations of these parameters.}
    \label{fig:maskclip_device_patterns}
\end{figure*}

\subsection{Device Implementation}

We developed MaskClip, a detachable sensing device designed to be both unobtrusive and functionally robust while maintaining the original comfort of the mask. As illustrated in Fig. \ref{fig:maskclip_device}, the hardware implementation consists of key components including the piezoelectric sensor, preamplifier circuit, codec, and ESP32 xiao microcontroller. Our primary design requirements were: (1) minimal visual footprint, (2) lightweight construction to prevent mask deformation, and (3) wireless connectivity to eliminate cable interference during daily use.
The core of MaskClip is built around the Seeed ESP32 microcontroller (S3 xiao), chosen for its compact form factor (approximately the size of a fingertip) and built-in wireless capabilities. This microcontroller features an integrated I2S (Inter-IC Sound) interface, which is crucial for high-quality digital audio acquisition. For analog-to-digital conversion, we employed the PCM1808PWR (Texas Instruments, USA), a high-precision 24-bit ADC optimized for audio applications.
The sensor assembly consists of a piezoelectric sensor operating at 3.3V, with its output signal conditioned through a custom amplification circuit before feeding into the ADC. We carefully tuned the amplification stage to match the ADC's input range while maintaining adequate signal-to-noise ratio for voice capture. The wireless transmission is handled by the nRF52840's integrated Bluetooth 5.0 module, which provides sufficient bandwidth for real-time audio streaming.
The entire circuit module is contained within a 25mm × 25mm form factor, with all components surface-mounted on a custom-designed PCB. Through careful component selection and layout optimization, we achieved a total weight of approximately 20g, including the housing. This compact design ensures that the device remains inconspicuous when attached to a face mask while maintaining stable contact with the mask surface for reliable sensing.
The power management system is designed to operate from a small lithium polymer battery, with careful consideration given to power consumption to enable extended use between charges. All electronic components are enclosed in a 3D-printed housing designed to provide adequate protection while maintaining minimal weight and size.
%




\subsection{Systematic Evaluation of Sensor Configuration Parameters}

To ensure robust performance across different usage scenarios, we conducted a systematic evaluation of four key parameters that could affect the sensor's performance.
Fig. \ref{fig:maskclip_device_patterns} provides an overview of our experimental conditions, showing the variations tested in clip materials (stainless steel versus plastic), sensor directions (inward-facing versus outward-facing), and positioning (non-woven mask versus fabric mask).

Positional sensitivity: We tested eight different positions (0mm to 70mm from the left edge, at 10mm intervals) to evaluate how sensor placement affects signal quality.
Sensor orientation: We compared two mounting orientations - inward-facing (toward the mask surface) and outward-facing (parallel to the mask surface) - to determine optimal sensor placement relative to the mask structure.
Mask material: We evaluated performance across two commonly used mask types: woven fabric masks and non-woven disposable masks, as material properties could affect vibration transmission.
Clip material: We compared stainless steel and plastic clips to assess how clip material properties influence acoustic transmission characteristics.

For all combinations of these parameters, we conducted acoustic measurements using a head and torso simulator (HATS) equipped with a standardized mouth simulator (deviation < 1dB across frequency bands). We used two voice conditions from the wTIMIT corpus: normal speech and whispered speech, using the phrase "this was easy for us" as our test stimulus. The HATS was positioned 1m above the floor and 1m from surrounding walls in an anechoic chamber (ambient noise: 28.8dB).
This comprehensive evaluation framework allows us to understand how each parameter independently and jointly affects sensor performance, enabling us to provide robust guidelines for real-world deployment.

\begin{figure*}[htbp]
    \centering
    \includegraphics[width=\linewidth,page=5]{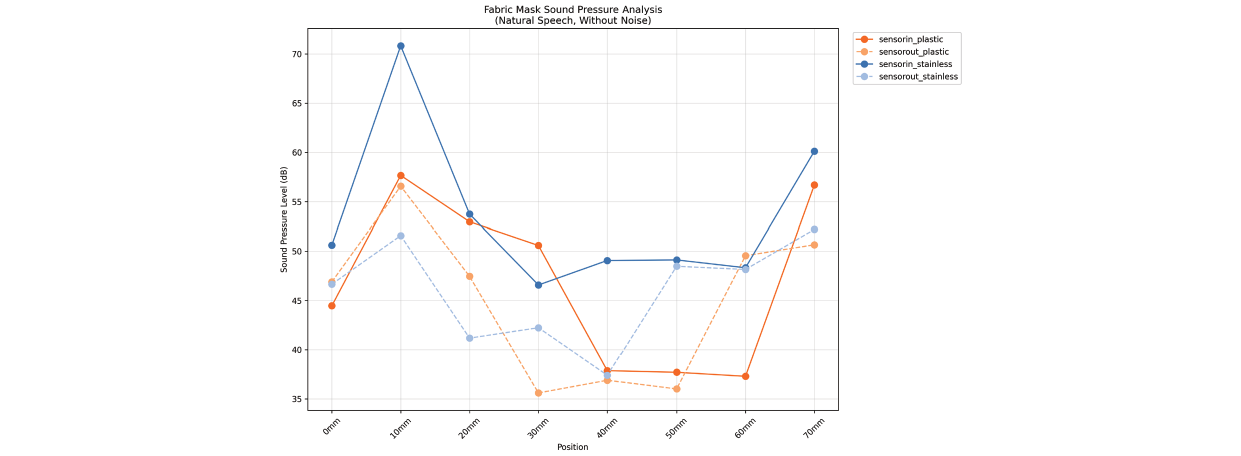}
    \caption{shows the results of testing different sensor configuration parameters for the MaskClip device. The graph compares performance between stainless steel and plastic clips, sensor orientations (inward-facing versus outward-facing), and positions ranging from 0mm to 70mm from the left edge of the mask. The stainless steel clip with an inward-facing sensor achieved the best performance, with peak performance of approximately 70 dB at the 10mm position. Even at distances of 30-60mm, the stainless steel configuration maintained stable signal detection around 45-50 dB. These results demonstrate that the stainless steel clip provides superior vibration transmission characteristics, making it the optimal choice for the final design.}
    \label{fig:eval_positions}
\end{figure*}
\subsection{Results: Influence of Position and Material Properties on Signal Detection}
The experimental results revealed several important characteristics about sensor positioning and materials that significantly impact the device's performance. 
As shown in Fig. \ref{fig:eval_positions}, our analysis of different configuration parameters demonstrated that the stainless steel clip with inward-facing sensor placement achieved optimal results, particularly at the 0mm and 10mm positions where the sensor makes contact with facial features like the cheeks and nose.
This is particularly evident with the stainless steel clip configuration (sensorin stainless), which achieved peak performance of approximately 70 dB at the 10mm position.
The comparison between sensor orientations (sensorin vs. sensorout) demonstrated that the inward-facing configuration consistently outperformed the outward-facing setup across all distances. This finding underscores the importance of mask thickness in signal transmission - when the sensor faces inward, it maintains closer contact with the mask surface, allowing for more effective capture of speech vibrations through the mask material.
A particularly noteworthy finding was the superior performance of the stainless steel clip configuration, especially at distances where direct skin contact was absent (around 30mm and beyond). While both plastic and stainless steel clips showed decreased performance at these distances, the stainless steel clip maintained relatively stable signal detection (around 45-50 dB) even at 30-60mm positions. This suggests that the material properties of stainless steel - including its rigidity and vibration transmission characteristics - contribute to more effective signal capture even when the sensor is not in direct contact with facial tissue.
Based on these comprehensive findings, we ultimately selected the inward-facing sensor configuration with a stainless steel clip (sensorin stainless) for our final design. This choice was driven by:

Superior overall performance across different distances
Consistent signal detection even without direct skin contact
Enhanced vibration transmission capabilities of the stainless steel material
Reliable performance in both contact and non-contact scenarios

\section{Evaluation}

\subsection{Hardware-based Noise Separation Performance}
\begin{figure*}[htbp]
    \centering
    \includegraphics[width=\linewidth, page=6]{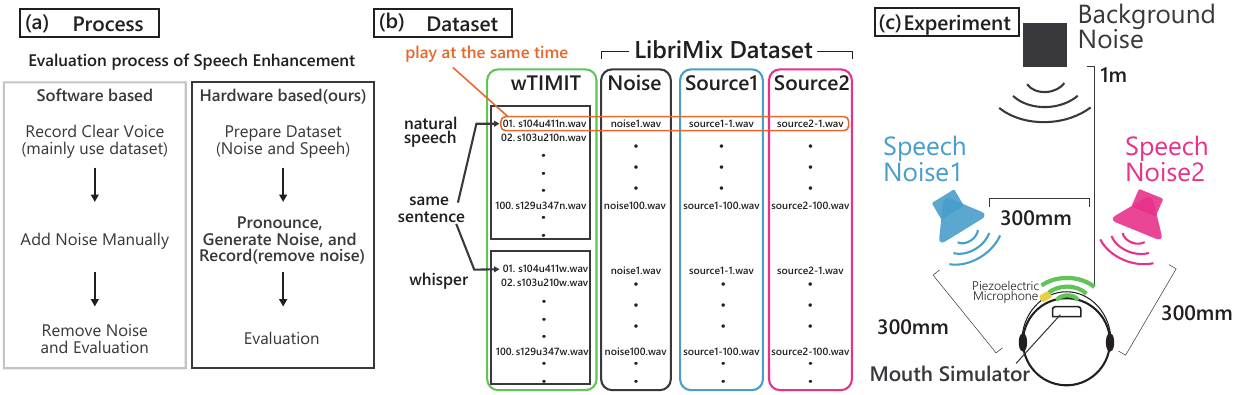}
    \caption{Comprehensive overview of the speech recognition evaluation process. (a) Comparison between software-based and hardware-based evaluation methodologies, (b) structure of the utilized datasets, and (c) experimental setup configuration. The evaluation employed wTIMIT and LibriMix datasets to simulate real-world conditions, enabling robust performance assessment.}
    \label{fig:evaluation_process}
\end{figure*}

While traditional software-based noise reduction methods are typically evaluated using clean speech recordings with artificially added noise, our hardware-based approach requires evaluation in real acoustic environments since the signal separation occurs at the point of capture (Fig. \ref{fig:evaluation_process} (a)). This fundamental difference in signal acquisition necessitates a comprehensive evaluation setup that reflects real-world usage conditions.
Common metrics for evaluating speech enhancement systems, such as Scale-Invariant Signal-to-Noise Ratio (SI-SNR), assume precise phase synchronization between the reference and enhanced signals. However, this assumption becomes problematic when working with speech recorded at typical deep learning sampling rates (16 kHz), where phase misalignment tolerance is limited to approximately 16 samples. Given these constraints, we chose to evaluate our system using Character Error Rate (CER) from general-purpose speech recognition AI systems, as this metric better reflects real-world usability and does not depend on phase alignment.

\subsection{Experimental Setup}
We constructed a controlled noise environment consisting of three primary components designed to simulate realistic usage scenarios:
Two speakers generating conversational interference (speech noise) were positioned 30cm from the subject, representing a challenging test condition that is closer than typical face-to-face conversation distances. This distance was deliberately chosen to evaluate performance under more demanding conditions, similar to what might be encountered in crowded medical settings or when colleagues are seated side-by-side. 
A third speaker generating background noise was positioned 1m away to reproduce ambient environmental noise. The test subject was simulated using a SAMAR4700M head and torso simulator (HATS) with mouth simulation capabilities, positioned 1m above the floor.
The sound pressure level was precisely calibrated to 60dB at 30cm using a precision sound level meter (measurement error < 0.1dB). All recordings were conducted in an anechoic chamber lined with porous materials, with an ambient noise level of 28.8dB, ensuring consistent and reproducible test conditions.

\subsection{Result of Speech Recognition}

\begin{figure*}[htbp]
    \centering
    \includegraphics[width=\linewidth, page=7]{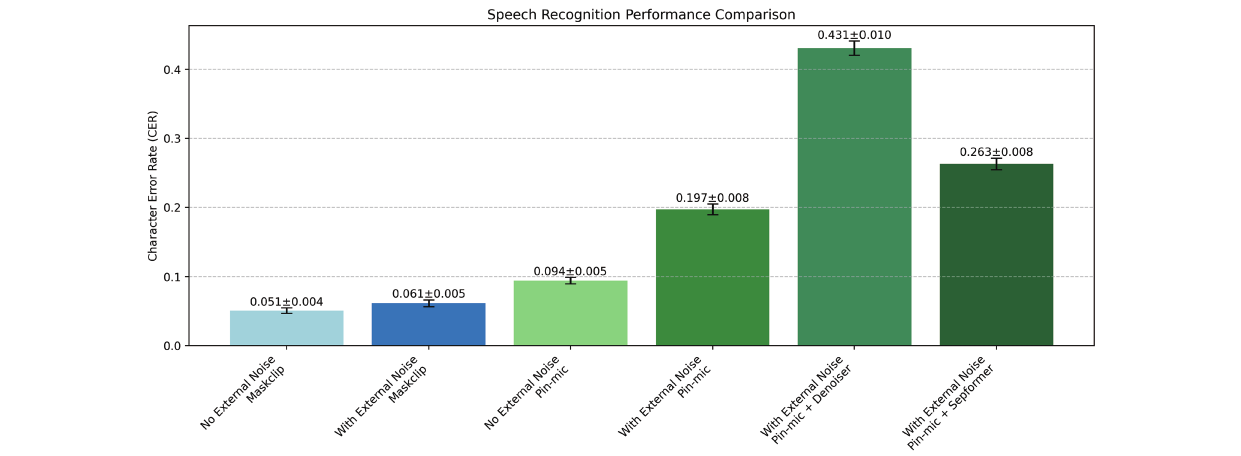}
    \caption{Final results of speech recognition evaluation. The graph compares Character Error Rates (CER) under various conditions (with and without noise, different processing methods). Results demonstrate MaskClip system's superior performance, particularly in noisy environments, compared to conventional approaches.}
    \label{fig:enter-label}
\end{figure*}

To evaluate the real-world performance of our system, we used OpenAI's Whisper-Large-V3~\cite{openai-whisper}, a state-of-the-art general-purpose speech recognition model. This model was specifically chosen for its robust performance in noisy environments and its ability to handle diverse acoustic conditions without requiring additional noise reduction preprocessing. Such characteristics make it particularly suitable for evaluating speech recognition performance across varying noise conditions.

We should clarify that in our evaluation setup, the Whisper-Large-V3 model was run on a separate GPU-equipped computer, not on the ESP32 xiao microcontroller itself. The ESP32 xiao was used solely for audio capture and wireless transmission of the recorded signals for subsequent processing.

For our evaluation dataset, we selected 1,640 utterances from the wTIMIT corpus~\cite{wtimit} where the same speaker recorded identical sentences in both normal and whispered speech. To create realistic noisy conditions, we used two types of noise from the WHAM! dataset~\cite{Wichern2019WHAM}: speech interference from two separate speakers and ambient background noise (Fig. \ref{fig:evaluation_process} (b)). The speech interference was played through two speakers positioned 30cm from the subject, while the background noise was played through a speaker at 1m distance. This paired dataset design allows for direct comparison of recognition performance between normal and whispered speech conditions under controlled variables, while the structured noise configuration simulates typical real-world environments where both conversational interference and ambient noise are present.

\begin{figure*}[htbp]
    \centering
    \includegraphics[width=\linewidth, page=8]{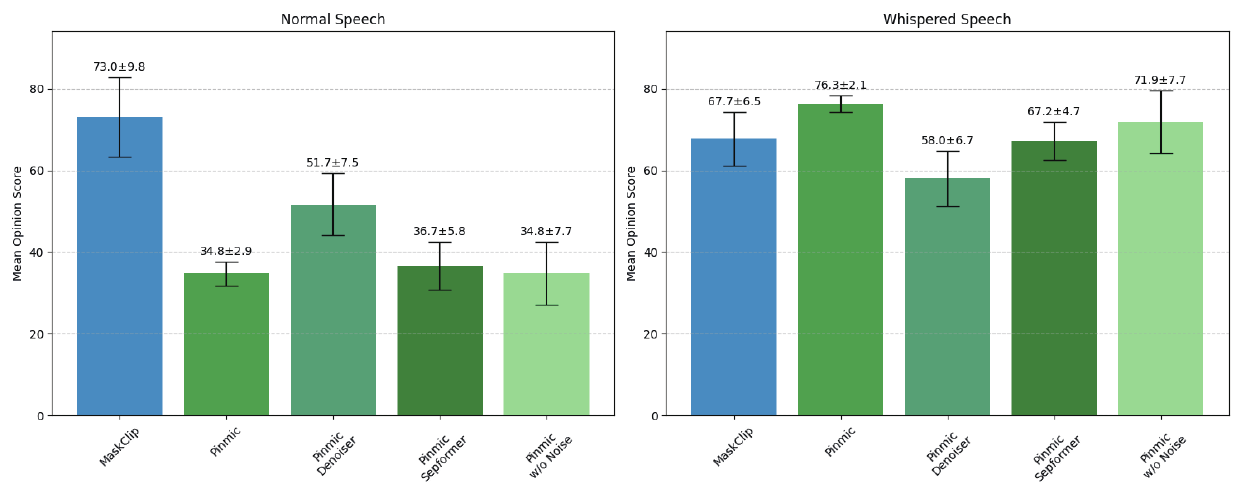}
    \caption{Subjective audio quality evaluation results for normal speech (left) and whispered speech (right) across different speech recognition systems. For normal speech, MaskClip achieved the highest score of 73.0±9.8 points, significantly outperforming the standard microphone which scored 34.8±2.9 points. Under whispered speech conditions, MaskClip maintained competitive performance with 67.7±6.5 points, demonstrating comparable or superior quality to other methods including the standard microphone (76.3±2.1 points), Denoiser (58.0±6.7 points), and Sepformer (67.2±4.7 points). Error bars indicate the consistency and reliability of these evaluations.}
    \label{fig:eval_mushra}
\end{figure*}

The evaluation encompasses three systems: the proposed MaskClip system, a conventional Pin-mic setup, and implementations with noise processing approaches (denoiser~\cite{denoiser} and sepformer~\cite{sepformer}). Both microphones were positioned 2cm from the mask's center on the opposite side, ensuring no direct contact with HATS. To simulate realistic noisy conditions, we supplemented the speech with interfering sounds from two speakers using the LibriMix dataset and background noise from the WHAM! dataset~\cite{Wichern2019WHAM} (Fig. \ref{fig:evaluation_process} (c))

The Character Error Rate (CER) was calculated by comparing the Whisper model's transcriptions against the ground truth texts. As shown in Fig. 7, the experimental data yielded the following performance metrics: The MaskClip system demonstrated CER values of 5.1\% without external noise and 6.1\% with external noise present. In comparison, the Pin-mic system recorded CER values of 9.4\% without external noise, increasing to 19.7\% with external noise present. When noise processing techniques were applied to the Pin-mic system, the CER increased to 43.1\% with the denoiser and 26.3\% with sepformer~\cite{sepformer}.

The introduction of external noise produced varying effects across the systems. The MaskCIP system exhibited relatively minor degradation, with approximately 1\% increase in CER. In contrast, the Pin-mic system showed more substantial performance deterioration, with an increase of approximately 10.3\% in CER.
The implementation of noise processing techniques resulted in increased CER compared to the baseline Pin-mic system, suggesting potential limitations in their effectiveness for this specific application.

\subsection{Evaluation of Mean Opinion Score}

To evaluate the subjective audio quality of our system, we conducted a MUSHRA (MUltiple Stimuli with Hidden Reference and Anchor) listening test. The evaluation involved 102 participants recruited through the online crowdsourcing platform Prolific, comprising 46 males, 54 females, and 1 participant who preferred not to disclose their gender. The mean age of participants was 32.3 years (SD = 10.3), and each participant received compensation of 7.5 euros.

\subsubsection{Evaluation Protocol}
The MUSHRA test was conducted using a web-based interface following the ITU-R BS.1534-3 recommendation. Participants used their own headphones and were instructed to adjust volume to a comfortable level before beginning the test. The interface presented all five conditions (standard microphone, MaskClip, standard microphone with Denoiser, standard microphone with Sepformer, and standard microphone with noise) simultaneously on a single screen for each audio sample, allowing direct comparison between methods.

The audio samples were pre-recorded from our experimental setup described in Section 4.2, where the HATS system was used to generate speech while each recording system (MaskClip and reference microphone) was positioned at the same 2cm distance from the mask center. For the software-enhanced conditions, the standard microphone recordings were processed using the respective algorithms. 

We used a scale from 0 to 100 as specified in the MUSHRA standard, where 0-20 represents 'bad' quality, 21-40 'poor', 41-60 'fair', 61-80 'good', and 81-100 'excellent'. This standardized scale was chosen to align with established methodologies in audio quality evaluation and to enable direct comparison with other studies. While free magnitude estimation offers advantages in avoiding scale constraints, the MUSHRA method was selected for its established validity in comparative audio quality assessment and its ability to reveal relative differences between systems.

\subsubsection{Randomization and Bias Control}
To minimize order effects and potential biases, both the presentation order of audio samples and the arrangement of methods on the interface were randomized for each participant. This randomization ensures that no systematic bias was introduced by a fixed evaluation sequence. Additionally, participants could listen to each sample as many times as needed and adjust their ratings until they were satisfied with their assessment across all conditions, thereby addressing potential memory limitations.

As shown in Fig. \ref{fig:eval_mushra}, for normal speech, MaskClip achieved the highest score of 73.0±9.8 points, significantly outperforming the standard microphone which scored 34.8±2.9 points. This difference was statistically significant as confirmed by paired t-tests (t = 3.564, p < 0.01). MaskClip also demonstrated superior performance compared to the Sepformer implementation (36.7±5.8 points), with a significant difference (t = 2.898, p < 0.05), establishing its advantage in normal speech conditions.

In whispered speech conditions, MaskClip maintained competitive performance with a score of 67.7±6.5 points, demonstrating comparable or superior quality to other methods including the standard microphone (76.3±2.1 points), Denoiser (58.0±6.7 points), and Sepformer (67.2±4.7 points). Notably, there was no significant difference between MaskClip and the noise-free standard microphone (71.9±7.7 points; t = -0.304, p = 0.761), indicating that MaskClip achieves audio quality comparable to conventional approaches for whispered speech.

These results demonstrate MaskClip's superior subjective audio quality for normal speech while maintaining performance parity with existing methods for whispered speech. This finding suggests that MaskClip offers consistent performance across various speaking conditions, representing a significant advancement in speech input technology. The standard errors indicated by error bars in Fig. \ref{fig:eval_mushra} confirm the consistency and reliability of these evaluations.

\section{Discussion}


In this research, we developed and validated MaskClip, a voice input device that utilizes mask surface vibrations to enable reliable speech recognition in challenging environments. Our evaluation results demonstrate practical value across various professional settings. Below, we discuss both the applications enabled by this technology and a detailed analysis of our evaluation results and their implications.

\subsection{Potential Application}
\begin{figure*}[htbp]
    \centering
    \includegraphics[width=\linewidth, page=9]{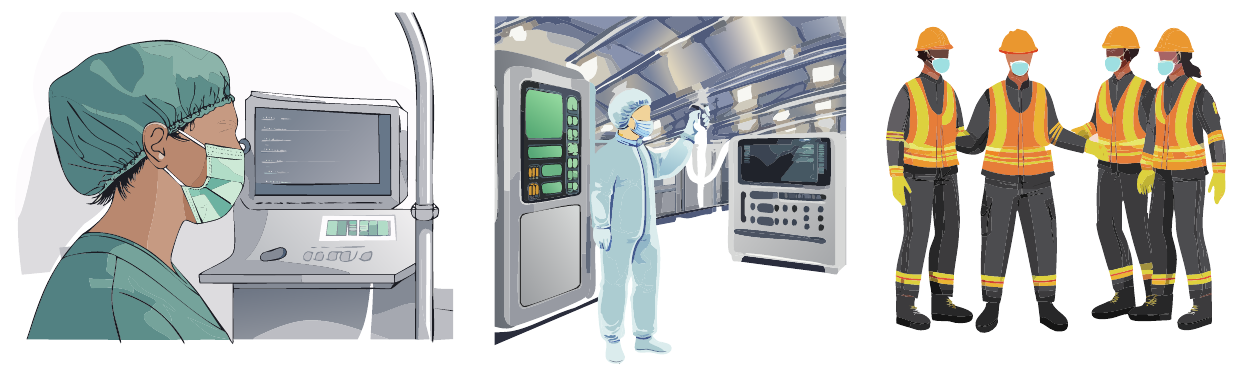}
    \caption{Illustration of MaskClip's potential applications across different environments. Left: A medical professional using voice commands to control equipment in an operating room while maintaining sterility. Center: A cleanroom technician performing quality control procedures with clear voice documentation capabilities. Right: Industrial workers coordinating tasks in a high-noise environment while maintaining respiratory protection. These scenarios demonstrate MaskClip's versatility in enabling clear voice communication while preserving protective mask functionality.}
    \label{fig:potential_application}
\end{figure*}

The MaskClip system demonstrates significant potential across various professional environments where clear voice communication while wearing masks is crucial. 
As illustrated in Fig. \ref{fig:potential_application}, these applications range from medical settings where surgeons need hands-free control of equipment, to cleanroom operations in semiconductor manufacturing facilities, and industrial environments where workers must maintain clear communication while wearing protective equipment.

In medical settings, the system enables surgeons and healthcare professionals to control medical equipment and documentation systems hands-free during procedures. This is particularly valuable in operating rooms where sterility must be maintained, allowing surgeons to adjust equipment settings, control visualization systems, and document procedures in real-time without breaking sterile technique. In infectious disease wards, the system facilitates clear communication between healthcare workers wearing N95 masks or other high-filtration masks, enhancing safety and efficiency during critical patient care scenarios.

In industrial applications, MaskClip provides valuable solutions for cleanroom operations in semiconductor manufacturing facilities and biomedical research laboratories. Workers can maintain cleanroom protocols while effectively logging data and controlling systems through voice commands, eliminating the need to remove protective equipment. The system's superior noise rejection capabilities make it particularly effective in high-noise industrial environments such as factory floors and construction sites, where traditional headset microphones often struggle with ambient noise interference.

The system also shows promise in emergency services and public safety applications. During disaster response operations, rescue workers wearing dust masks can maintain clear communication in challenging environments filled with smoke, dust, or debris. This capability extends to various public safety scenarios where personnel must coordinate response efforts while maintaining respiratory protection. The system's ability to provide clear voice communication without compromising protective equipment makes it particularly valuable in these critical scenarios.

\subsection{Physical Mechanisms of Vibration Detection and Performance Characteristics}

One significant finding of this research is that direct skin contact has minimal impact on voice detection, which can be explained by the vibration propagation mechanism through the mask surface. Our experimental results showed peak performance of approximately 70 dB at the 10mm position with the stainless steel clip, maintaining stable signal detection of 45-50 dB even at positions between 30-60mm. These findings demonstrate that stable voice detection is possible without relying on skin contact, unlike traditional NAM microphones or throat microphones.
However, further verification is needed regarding performance under dynamic usage conditions, such as walking or sudden head movements. Our experimental results showed that in environments with external noise, the Character Error Rate (CER) was approximately 6.1\% for the MaskClip system compared to 19.7\% for the conventional Pin-mic system. This substantial difference suggests that our system's fundamental principle of detecting mask surface vibrations provides inherent resistance to environmental noise.

\subsection{Advancement Toward Multimodal Voice Processing}
The experimental results suggest possibilities for further performance improvements. Notably, MaskClip demonstrated significantly superior performance compared to conventional noise processing technologies, which showed CER values of 43.1\% for Denoiser and 26.3\% for Sepformer. These figures indicate potential for further enhancement by combining physical voice detection mechanisms with software-based signal processing.
Given that MaskClip achieved CER values of 5.1\% in normal speech and 6.1\% in noisy environments, integration with Voice Activity Detection (VAD) could lead to further accuracy improvements. Development from the current single-sensor configuration to a multi-sensor spatial noise separation approach could potentially improve current CER values. Additionally, the observed CER degradation of approximately 1\% in noisy environments could be further reduced through adaptive signal processing.

\subsection{System Integration and Application Opportunities}
Our experimental results suggest that the system possesses sufficient performance capabilities for practical use in medical settings. Particularly, the minimal CER increase of approximately 1\% in noisy environments supports its viability across various acoustic environments in medical settings. In contrast, the approximately 10.3\% CER increase observed in the Pin-mic system clearly demonstrates the limitations of conventional technology.
Based on these findings, the system shows particular promise for applications in medical voice interfaces, voice commands in clean rooms, and general speech recognition tasks in noisy environments. This research significantly advances the feasibility of voice interfaces in medical settings. Future developments will focus on long-term evaluation in diverse real-world environments and integration with multimodal signal processing for further performance enhancements.
Our findings not only validate the technical viability of MaskClip but also highlight its potential as a practical solution for voice interaction in challenging acoustic environments. The consistent performance across various conditions, particularly the minimal degradation in noisy environments, positions this technology as a promising advancement in speech input systems.

\section{Conclusion}

In this paper, we introduced a novel principle for speech input in mask-required environments such as medical facilities and cleanrooms by detecting the vibrations of mask surfaces. Our approach utilizes a detachable piezoelectric sensor mounted on the mask surface to selectively capture vibrations transmitted through the mask during speech, enabling noise-robust voice input. While conventional throat microphones~\cite{throatmic1} and NAM microphones~\cite{nam_microphone} require direct skin contact, our method potentially offers reduced discomfort and motion artifacts through its non-contact sensing principle.

Our evaluation demonstrated that the system achieved a Character Error Rate (CER) of 5.1\% in quiet conditions and 6.1\% in noisy environments. These results show improvement compared to a conventional pin microphone setup, which exhibited CERs of 9.4\% and 19.7\% respectively. The system's hardware-based noise suppression approach eliminates the need for GPU-based processing, offering practical advantages for real-world deployment.

The proposed method has potential applications in various professional settings where masks are mandatory, including voice control of medical equipment, documentation in cleanrooms, and communication in industrial environments. By implementing noise suppression through hardware rather than computational processing, the system offers a practical solution for voice interfaces in these settings.

Future work should address several key areas: performance verification during walking and head movements, enhancement of spatial noise separation through multi-sensor configurations, and potential improvements in recognition accuracy through integration with Voice Activity Detection (VAD). These developments could further enhance the system's capabilities and broaden its practical applications.

\section*{acknowledgement}
 This work was supported by JST ACT-X Grant JPMJAX23KG, JST Moonshot R\&D Grant JPMJMS2012, and JPNP23025 commissioned by the New Energy and Industrial Technology Development Organization (NEDO).

\bibliographystyle{ACM-Reference-Format}
\bibliography{main}


\begin{thebibliography}{55}


\ifx \showCODEN    \undefined \def \showCODEN     #1{\unskip}     \fi
\ifx \showDOI      \undefined \def \showDOI       #1{#1}\fi
\ifx \showISBNx    \undefined \def \showISBNx     #1{\unskip}     \fi
\ifx \showISBNxiii \undefined \def \showISBNxiii  #1{\unskip}     \fi
\ifx \showISSN     \undefined \def \showISSN      #1{\unskip}     \fi
\ifx \showLCCN     \undefined \def \showLCCN      #1{\unskip}     \fi
\ifx \shownote     \undefined \def \shownote      #1{#1}          \fi
\ifx \showarticletitle \undefined \def \showarticletitle #1{#1}   \fi
\ifx \showURL      \undefined \def \showURL       {\relax}        \fi
\providecommand\bibfield[2]{#2}
\providecommand\bibinfo[2]{#2}
\providecommand\natexlab[1]{#1}
\providecommand\showeprint[2][]{arXiv:#2}

\bibitem[Al-Karawi(2024)]%
        {Al-Karawi2023-zm}
\bibfield{author}{\bibinfo{person}{Khamis~A. Al-Karawi}.} \bibinfo{year}{2024}\natexlab{}.
\newblock \showarticletitle{Face mask effects on speaker verification performance in the presence of noise}.
\newblock \bibinfo{journal}{\emph{Multimedia Tools and Applications}} \bibinfo{volume}{83}, \bibinfo{number}{2} (\bibinfo{year}{2024}), \bibinfo{pages}{4811--4824}.
\newblock
\showISSN{1573-7721}
\urldef\tempurl%
\url{https://doi.org/10.1007/s11042-023-15824-w}
\showDOI{\tempurl}


\bibitem[Bauer(1962)]%
        {A-century-of-microphones}
\bibfield{author}{\bibinfo{person}{B.~B. Bauer}.} \bibinfo{year}{1962}\natexlab{}.
\newblock \showarticletitle{A Century of Microphones}.
\newblock \bibinfo{journal}{\emph{Proceedings of the IRE}} \bibinfo{volume}{50}, \bibinfo{number}{5} (\bibinfo{year}{1962}), \bibinfo{pages}{719--729}.
\newblock
\urldef\tempurl%
\url{https://doi.org/10.1109/JRPROC.1962.288106}
\showDOI{\tempurl}


\bibitem[Beach et~al\mbox{.}(2019)]%
        {SleepMask}
\bibfield{author}{\bibinfo{person}{Christopher Beach}, \bibinfo{person}{Nazmul Karim}, {and} \bibinfo{person}{Alexander~J. Casson}.} \bibinfo{year}{2019}\natexlab{}.
\newblock \showarticletitle{A Graphene-Based Sleep Mask for Comfortable Wearable Eye Tracking}. In \bibinfo{booktitle}{\emph{2019 41st Annual International Conference of the IEEE Engineering in Medicine and Biology Society (EMBC)}}. \bibinfo{pages}{6693--6696}.
\newblock
\urldef\tempurl%
\url{https://doi.org/10.1109/EMBC.2019.8857198}
\showDOI{\tempurl}


\bibitem[Chatterjee et~al\mbox{.}(2022)]%
        {clearbuds}
\bibfield{author}{\bibinfo{person}{Ishan Chatterjee}, \bibinfo{person}{Maruchi Kim}, \bibinfo{person}{Vivek Jayaram}, \bibinfo{person}{Shyamnath Gollakota}, \bibinfo{person}{Ira Kemelmacher}, \bibinfo{person}{Shwetak Patel}, {and} \bibinfo{person}{Steven~M. Seitz}.} \bibinfo{year}{2022}\natexlab{}.
\newblock \showarticletitle{ClearBuds: Wireless Binaural Earbuds for Learning-Based Speech Enhancement}. In \bibinfo{booktitle}{\emph{Proceedings of the 20th Annual International Conference on Mobile Systems, Applications and Services}} (Portland, Oregon) \emph{(\bibinfo{series}{MobiSys '22})}. \bibinfo{publisher}{Association for Computing Machinery}, \bibinfo{address}{New York, NY, USA}, \bibinfo{pages}{384–396}.
\newblock
\showISBNx{9781450391856}
\urldef\tempurl%
\url{https://doi.org/10.1145/3498361.3538933}
\showDOI{\tempurl}


\bibitem[Ding et~al\mbox{.}(2022)]%
        {UltraSpeech}
\bibfield{author}{\bibinfo{person}{Han Ding}, \bibinfo{person}{Yizhan Wang}, \bibinfo{person}{Hao Li}, \bibinfo{person}{Cui Zhao}, \bibinfo{person}{Ge Wang}, \bibinfo{person}{Wei Xi}, {and} \bibinfo{person}{Jizhong Zhao}.} \bibinfo{year}{2022}\natexlab{}.
\newblock \showarticletitle{UltraSpeech: Speech Enhancement by Interaction between Ultrasound and Speech}.
\newblock \bibinfo{journal}{\emph{Proc. ACM Interact. Mob. Wearable Ubiquitous Technol.}} \bibinfo{volume}{6}, \bibinfo{number}{3}, Article \bibinfo{articleno}{111} (\bibinfo{date}{Sept.} \bibinfo{year}{2022}), \bibinfo{numpages}{25}~pages.
\newblock
\urldef\tempurl%
\url{https://doi.org/10.1145/3550303}
\showDOI{\tempurl}


\bibitem[Dong et~al\mbox{.}(2024)]%
        {ReHEarSSE}
\bibfield{author}{\bibinfo{person}{Xuefu Dong}, \bibinfo{person}{Yifei Chen}, \bibinfo{person}{Yuuki Nishiyama}, \bibinfo{person}{Kaoru Sezaki}, \bibinfo{person}{Yuntao Wang}, \bibinfo{person}{Ken Christofferson}, {and} \bibinfo{person}{Alex Mariakakis}.} \bibinfo{year}{2024}\natexlab{}.
\newblock \showarticletitle{ReHEarSSE: Recognizing Hidden-in-the Ear Silently Spelled Expressions}. In \bibinfo{booktitle}{\emph{Proceedings of the CHI Conference on Human Factors in Computing Systems}} (Honolulu, HI, USA) \emph{(\bibinfo{series}{CHI '24})}. \bibinfo{publisher}{Association for Computing Machinery}, \bibinfo{address}{New York, NY, USA}, Article \bibinfo{articleno}{321}, \bibinfo{numpages}{16}~pages.
\newblock
\showISBNx{9798400703300}
\urldef\tempurl%
\url{https://doi.org/10.1145/3613904.3642095}
\showDOI{\tempurl}


\bibitem[Duan et~al\mbox{.}(2024)]%
        {earse}
\bibfield{author}{\bibinfo{person}{Di Duan}, \bibinfo{person}{Yongliang Chen}, \bibinfo{person}{Weitao Xu}, {and} \bibinfo{person}{Tianxing Li}.} \bibinfo{year}{2024}\natexlab{}.
\newblock \showarticletitle{EarSE: Bringing Robust Speech Enhancement to COTS Headphones}.
\newblock \bibinfo{journal}{\emph{Proc. ACM Interact. Mob. Wearable Ubiquitous Technol.}} \bibinfo{volume}{7}, \bibinfo{number}{4}, Article \bibinfo{articleno}{158} (\bibinfo{date}{Jan.} \bibinfo{year}{2024}), \bibinfo{numpages}{33}~pages.
\newblock
\urldef\tempurl%
\url{https://doi.org/10.1145/3631447}
\showDOI{\tempurl}


\bibitem[Défossez et~al\mbox{.}(2020)]%
        {denoiser}
\bibfield{author}{\bibinfo{person}{Alexandre Défossez}, \bibinfo{person}{Gabriel Synnaeve}, {and} \bibinfo{person}{Yossi Adi}.} \bibinfo{year}{2020}\natexlab{}.
\newblock \showarticletitle{{Real Time Speech Enhancement in the Waveform Domain}}. In \bibinfo{booktitle}{\emph{Proc. Interspeech 2020}}. \bibinfo{pages}{3291--3295}.
\newblock
\urldef\tempurl%
\url{https://doi.org/10.21437/Interspeech.2020-2409}
\showDOI{\tempurl}


\bibitem[Erzin(2009)]%
        {Erzin2009-pd}
\bibfield{author}{\bibinfo{person}{Engin Erzin}.} \bibinfo{year}{2009}\natexlab{}.
\newblock \showarticletitle{Improving Throat Microphone Speech Recognition by Joint Analysis of Throat and Acoustic Microphone Recordings}.
\newblock \bibinfo{journal}{\emph{IEEE Transactions on Audio, Speech, and Language Processing}} \bibinfo{volume}{17}, \bibinfo{number}{7} (\bibinfo{year}{2009}), \bibinfo{pages}{1316--1324}.
\newblock
\urldef\tempurl%
\url{https://doi.org/10.1109/TASL.2009.2016733}
\showDOI{\tempurl}


\bibitem[Fukumoto(2018)]%
        {fukumoto}
\bibfield{author}{\bibinfo{person}{Masaaki Fukumoto}.} \bibinfo{year}{2018}\natexlab{}.
\newblock \showarticletitle{SilentVoice: Unnoticeable Voice Input by Ingressive Speech}. In \bibinfo{booktitle}{\emph{Proceedings of the 31st Annual ACM Symposium on User Interface Software and Technology}} (Berlin, Germany) \emph{(\bibinfo{series}{UIST '18})}. \bibinfo{publisher}{Association for Computing Machinery}, \bibinfo{address}{New York, NY, USA}, \bibinfo{pages}{237–246}.
\newblock
\showISBNx{9781450359481}
\urldef\tempurl%
\url{https://doi.org/10.1145/3242587.3242603}
\showDOI{\tempurl}


\bibitem[Gonzalez et~al\mbox{.}(2017)]%
        {DirectSpeechReconstruction}
\bibfield{author}{\bibinfo{person}{Jose~A. Gonzalez}, \bibinfo{person}{Lam~A. Cheah}, \bibinfo{person}{Angel~M. Gomez}, \bibinfo{person}{Phil~D. Green}, \bibinfo{person}{James~M. Gilbert}, \bibinfo{person}{Stephen~R. Ell}, \bibinfo{person}{Roger~K. Moore}, {and} \bibinfo{person}{Ed Holdsworth}.} \bibinfo{year}{2017}\natexlab{}.
\newblock \showarticletitle{Direct Speech Reconstruction From Articulatory Sensor Data by Machine Learning}.
\newblock \bibinfo{journal}{\emph{IEEE/ACM Transactions on Audio, Speech, and Language Processing}} \bibinfo{volume}{25}, \bibinfo{number}{12} (\bibinfo{year}{2017}), \bibinfo{pages}{2362--2374}.
\newblock
\urldef\tempurl%
\url{https://doi.org/10.1109/TASLP.2017.2757263}
\showDOI{\tempurl}


\bibitem[Guo and Liang(2023)]%
        {TexonMask}
\bibfield{author}{\bibinfo{person}{Zengrong Guo} {and} \bibinfo{person}{Rong-Hao Liang}.} \bibinfo{year}{2023}\natexlab{}.
\newblock \showarticletitle{TexonMask: Facial Expression Recognition Using Textile Electrodes on Commodity Facemasks}. In \bibinfo{booktitle}{\emph{Proceedings of the 2023 CHI Conference on Human Factors in Computing Systems}} (Hamburg, Germany) \emph{(\bibinfo{series}{CHI '23})}. \bibinfo{publisher}{Association for Computing Machinery}, \bibinfo{address}{New York, NY, USA}, Article \bibinfo{articleno}{627}, \bibinfo{numpages}{15}~pages.
\newblock
\showISBNx{9781450394215}
\urldef\tempurl%
\url{https://doi.org/10.1145/3544548.3581295}
\showDOI{\tempurl}


\bibitem[Hirahara et~al\mbox{.}(2010)]%
        {NAM2}
\bibfield{author}{\bibinfo{person}{Tatsuya Hirahara}, \bibinfo{person}{Makoto Otani}, \bibinfo{person}{Shota Shimizu}, \bibinfo{person}{Tomoki Toda}, \bibinfo{person}{Keigo Nakamura}, \bibinfo{person}{Yoshitaka Nakajima}, {and} \bibinfo{person}{Kiyohiro Shikano}.} \bibinfo{year}{2010}\natexlab{}.
\newblock \showarticletitle{Silent-speech enhancement using body-conducted vocal-tract resonance signals}.
\newblock \bibinfo{journal}{\emph{Speech Communication}} \bibinfo{volume}{52}, \bibinfo{number}{4} (\bibinfo{year}{2010}), \bibinfo{pages}{301--313}.
\newblock
\showISSN{0167-6393}
\urldef\tempurl%
\url{https://doi.org/10.1016/j.specom.2009.12.001}
\showDOI{\tempurl}
\newblock
\shownote{Silent Speech Interfaces}.


\bibitem[Hiraki et~al\mbox{.}(2023)]%
        {whispermask_chiEA}
\bibfield{author}{\bibinfo{person}{Hirotaka Hiraki}, \bibinfo{person}{Shusuke Kanazawa}, \bibinfo{person}{Takahiro Miura}, \bibinfo{person}{Manabu Yoshida}, \bibinfo{person}{Masaaki Mochimaru}, {and} \bibinfo{person}{Jun Rekimoto}.} \bibinfo{year}{2023}\natexlab{}.
\newblock \showarticletitle{External noise reduction using WhisperMask, a mask-type wearable microphone}. In \bibinfo{booktitle}{\emph{Extended Abstracts of the 2023 CHI Conference on Human Factors in Computing Systems}} (Hamburg, Germany) \emph{(\bibinfo{series}{CHI EA '23})}. \bibinfo{publisher}{Association for Computing Machinery}, \bibinfo{address}{New York, NY, USA}, Article \bibinfo{articleno}{454}, \bibinfo{numpages}{5}~pages.
\newblock
\showISBNx{9781450394222}
\urldef\tempurl%
\url{https://doi.org/10.1145/3544549.3583936}
\showDOI{\tempurl}


\bibitem[Hiraki et~al\mbox{.}(2024)]%
        {whispermask_ahs}
\bibfield{author}{\bibinfo{person}{Hirotaka Hiraki}, \bibinfo{person}{Shusuke Kanazawa}, \bibinfo{person}{Takahiro Miura}, \bibinfo{person}{Manabu Yoshida}, \bibinfo{person}{Masaaki Mochimaru}, {and} \bibinfo{person}{Jun Rekimoto}.} \bibinfo{year}{2024}\natexlab{}.
\newblock \showarticletitle{WhisperMask: a noise suppressive mask-type microphone for whisper speech}. In \bibinfo{booktitle}{\emph{Proceedings of the Augmented Humans International Conference 2024}} (Melbourne, VIC, Australia) \emph{(\bibinfo{series}{AHs '24})}. \bibinfo{publisher}{Association for Computing Machinery}, \bibinfo{address}{New York, NY, USA}, \bibinfo{pages}{1–14}.
\newblock
\showISBNx{9798400709807}
\urldef\tempurl%
\url{https://doi.org/10.1145/3652920.3652925}
\showDOI{\tempurl}


\bibitem[Hiraki and Rekimoto(2021)]%
        {SilentMask}
\bibfield{author}{\bibinfo{person}{Hirotaka Hiraki} {and} \bibinfo{person}{Jun Rekimoto}.} \bibinfo{year}{2021}\natexlab{}.
\newblock \showarticletitle{SilentMask: Mask-Type Silent Speech Interface with Measurement of Mouth Movement}. In \bibinfo{booktitle}{\emph{Augmented Humans Conference 2021}} (Rovaniemi, Finland) \emph{(\bibinfo{series}{AHs'21})}. \bibinfo{publisher}{Association for Computing Machinery}, \bibinfo{address}{New York, NY, USA}, \bibinfo{pages}{86–90}.
\newblock
\showISBNx{9781450384285}
\urldef\tempurl%
\url{https://doi.org/10.1145/3458709.3458985}
\showDOI{\tempurl}


\bibitem[Hu et~al\mbox{.}(2024)]%
        {robustGER}
\bibfield{author}{\bibinfo{person}{Yuchen Hu}, \bibinfo{person}{Chen Chen}, \bibinfo{person}{Chao{-}Han~Huck Yang}, \bibinfo{person}{Ruizhe Li}, \bibinfo{person}{Chao Zhang}, \bibinfo{person}{Pin{-}Yu Chen}, {and} \bibinfo{person}{Engsiong Chng}.} \bibinfo{year}{2024}\natexlab{}.
\newblock \showarticletitle{Large Language Models are Efficient Learners of Noise-Robust Speech Recognition}. In \bibinfo{booktitle}{\emph{The Twelfth International Conference on Learning Representations, {ICLR} 2024, Vienna, Austria, May 7-11, 2024}}. \bibinfo{publisher}{OpenReview.net}.
\newblock
\urldef\tempurl%
\url{https://doi.org/10.48550/arXiv.2401.10446}
\showDOI{\tempurl}


\bibitem[Ingalls(1987)]%
        {Throat-microphone}
\bibfield{author}{\bibinfo{person}{Robert Ingalls}.} \bibinfo{year}{1987}\natexlab{}.
\newblock \showarticletitle{{Throat microphone}}.
\newblock \bibinfo{journal}{\emph{The Journal of the Acoustical Society of America}} \bibinfo{volume}{81}, \bibinfo{number}{3} (\bibinfo{date}{03} \bibinfo{year}{1987}), \bibinfo{pages}{809--809}.
\newblock
\showISSN{0001-4966}
\urldef\tempurl%
\url{https://doi.org/10.1121/1.394659}
\showDOI{\tempurl}
\showeprint{https://pubs.aip.org/asa/jasa/article-pdf/81/3/809/12095011/809\_1\_online.pdf}


\bibitem[Kawaguchi and Matsumoto(2022)]%
        {throatmic2}
\bibfield{author}{\bibinfo{person}{Junki Kawaguchi} {and} \bibinfo{person}{Mitsuharu Matsumoto}.} \bibinfo{year}{2022}\natexlab{}.
\newblock \showarticletitle{Noise Reduction Combining a General Microphone and a Throat Microphone}.
\newblock \bibinfo{journal}{\emph{Sensors}} \bibinfo{volume}{22}, \bibinfo{number}{12} (\bibinfo{year}{2022}).
\newblock
\showISSN{1424-8220}
\urldef\tempurl%
\url{https://doi.org/10.3390/s2212 s4473}
\showDOI{\tempurl}


\bibitem[Khanna et~al\mbox{.}(2021)]%
        {jawsense}
\bibfield{author}{\bibinfo{person}{Prerna Khanna}, \bibinfo{person}{Tanmay Srivastava}, \bibinfo{person}{Shijia Pan}, \bibinfo{person}{Shubham Jain}, {and} \bibinfo{person}{Phuc Nguyen}.} \bibinfo{year}{2021}\natexlab{}.
\newblock \showarticletitle{JawSense: Recognizing Unvoiced Sound Using a Low-Cost Ear-Worn System}. In \bibinfo{booktitle}{\emph{Proceedings of the 22nd International Workshop on Mobile Computing Systems and Applications}} (Virtual, United Kingdom) \emph{(\bibinfo{series}{HotMobile '21})}. \bibinfo{publisher}{Association for Computing Machinery}, \bibinfo{address}{New York, NY, USA}, \bibinfo{pages}{44–49}.
\newblock
\showISBNx{9781450383233}
\urldef\tempurl%
\url{https://doi.org/10.1145/3446382.3448363}
\showDOI{\tempurl}


\bibitem[Kumazaki and Inoue(2020)]%
        {TransEmotion}
\bibfield{author}{\bibinfo{person}{Ryoga Kumazaki} {and} \bibinfo{person}{Akifumi Inoue}.} \bibinfo{year}{2020}\natexlab{}.
\newblock \showarticletitle{Development and Evaluation of a Mask-Type Display Transforming the Wearer's Impression}. In \bibinfo{booktitle}{\emph{Proceedings of 31st Australian Conference on Human-Computer-Interaction}} (Fremantle, WA, Australia) \emph{(\bibinfo{series}{OzCHI '19})}. \bibinfo{publisher}{Association for Computing Machinery}, \bibinfo{address}{New York, NY, USA}, \bibinfo{pages}{568–571}.
\newblock
\showISBNx{9781450376969}
\urldef\tempurl%
\url{https://doi.org/10.1145/3369457.3369533}
\showDOI{\tempurl}


\bibitem[Kunimi et~al\mbox{.}(2022)]%
        {E-mask}
\bibfield{author}{\bibinfo{person}{Yusuke Kunimi}, \bibinfo{person}{Masa Ogata}, \bibinfo{person}{Hirotaka Hiraki}, \bibinfo{person}{Motoshi Itagaki}, \bibinfo{person}{Shusuke Kanazawa}, {and} \bibinfo{person}{Masaaki Mochimaru}.} \bibinfo{year}{2022}\natexlab{}.
\newblock \showarticletitle{E-MASK: A Mask-Shaped Interface for Silent Speech Interaction with Flexible Strain Sensors}. In \bibinfo{booktitle}{\emph{Augmented Humans 2022}} (Kashiwa, Chiba, Japan) \emph{(\bibinfo{series}{AHs 2022})}. \bibinfo{publisher}{Association for Computing Machinery}, \bibinfo{address}{New York, NY, USA}, \bibinfo{pages}{26–34}.
\newblock
\showISBNx{9781450396325}
\urldef\tempurl%
\url{https://doi.org/10.1145/3519391.3519399}
\showDOI{\tempurl}


\bibitem[Kusabuka and Indo(2020)]%
        {ibuki}
\bibfield{author}{\bibinfo{person}{Takahiro Kusabuka} {and} \bibinfo{person}{Takuya Indo}.} \bibinfo{year}{2020}\natexlab{}.
\newblock \showarticletitle{IBUKI: Gesture Input Method Based on Breathing}. In \bibinfo{booktitle}{\emph{Adjunct Publication of the 33rd Annual ACM Symposium on User Interface Software and Technology}} (Virtual Event, USA) \emph{(\bibinfo{series}{UIST '20 Adjunct})}. \bibinfo{publisher}{Association for Computing Machinery}, \bibinfo{address}{New York, NY, USA}, \bibinfo{pages}{102–104}.
\newblock
\showISBNx{9781450375153}
\urldef\tempurl%
\url{https://doi.org/10.1145/3379350.3416134}
\showDOI{\tempurl}


\bibitem[Latif et~al\mbox{.}(2021)]%
        {Latif2021-wt}
\bibfield{author}{\bibinfo{person}{Siddique Latif}, \bibinfo{person}{Junaid Qadir}, \bibinfo{person}{Adnan Qayyum}, \bibinfo{person}{Muhammad Usama}, {and} \bibinfo{person}{Shahzad Younis}.} \bibinfo{year}{2021}\natexlab{}.
\newblock \showarticletitle{Speech Technology for Healthcare: Opportunities, Challenges, and State of the Art}.
\newblock \bibinfo{journal}{\emph{IEEE Reviews in Biomedical Engineering}}  \bibinfo{volume}{14} (\bibinfo{year}{2021}), \bibinfo{pages}{342--356}.
\newblock
\urldef\tempurl%
\url{https://doi.org/10.1109/RBME.2020.3006860}
\showDOI{\tempurl}


\bibitem[Lee et~al\mbox{.}(2020)]%
        {mascreen}
\bibfield{author}{\bibinfo{person}{Hyein Lee}, \bibinfo{person}{Yoonji Kim}, {and} \bibinfo{person}{Andrea Bianchi}.} \bibinfo{year}{2020}\natexlab{}.
\newblock \showarticletitle{MAScreen: Augmenting Speech with Visual Cues of Lip Motions, Facial Expressions, and Text Using a Wearable Display}. In \bibinfo{booktitle}{\emph{SIGGRAPH Asia 2020 Emerging Technologies}} (Virtual Event, Republic of Korea) \emph{(\bibinfo{series}{SA '20})}. \bibinfo{publisher}{Association for Computing Machinery}, \bibinfo{address}{New York, NY, USA}, Article \bibinfo{articleno}{2}, \bibinfo{numpages}{2}~pages.
\newblock
\showISBNx{9781450381109}
\urldef\tempurl%
\url{https://doi.org/10.1145/3415255.3422886}
\showDOI{\tempurl}


\bibitem[Lee(1998)]%
        {ICA_bss}
\bibfield{author}{\bibinfo{person}{Te-Won Lee}.} \bibinfo{year}{1998}\natexlab{}.
\newblock \bibinfo{booktitle}{\emph{Independent Component Analysis}}.
\newblock \bibinfo{publisher}{Springer US}, \bibinfo{address}{Boston, MA}, \bibinfo{pages}{27--66}.
\newblock
\showISBNx{978-1-4757-2851-4}
\urldef\tempurl%
\url{https://doi.org/10.1007/978-1-4757-2851-4_2}
\showDOI{\tempurl}


\bibitem[Lim(2010)]%
        {wtimit}
\bibfield{author}{\bibinfo{person}{Boon~Pang Lim}.} \bibinfo{year}{2010}\natexlab{}.
\newblock \showarticletitle{Computational differences between whispered and non-whispered speech}. \bibinfo{publisher}{PhD Thesis UIUC}.
\newblock


\bibitem[Luo and Mesgarani(2019)]%
        {Conv-tasnet}
\bibfield{author}{\bibinfo{person}{Yi Luo} {and} \bibinfo{person}{Nima Mesgarani}.} \bibinfo{year}{2019}\natexlab{}.
\newblock \showarticletitle{Conv-TasNet: Surpassing Ideal Time–Frequency Magnitude Masking for Speech Separation}.
\newblock \bibinfo{journal}{\emph{IEEE/ACM Transactions on Audio, Speech, and Language Processing}} \bibinfo{volume}{27}, \bibinfo{number}{8} (\bibinfo{year}{2019}), \bibinfo{pages}{1256--1266}.
\newblock
\urldef\tempurl%
\url{https://doi.org/10.1109/TASLP.2019.2915167}
\showDOI{\tempurl}


\bibitem[Mirzal(2017)]%
        {nmf_for_bss}
\bibfield{author}{\bibinfo{person}{Andri Mirzal}.} \bibinfo{year}{2017}\natexlab{}.
\newblock \showarticletitle{NMF versus ICA for blind source separation}.
\newblock \bibinfo{journal}{\emph{Advances in Data Analysis and Classification}} \bibinfo{volume}{11}, \bibinfo{number}{1} (\bibinfo{year}{2017}), \bibinfo{pages}{25--48}.
\newblock
\urldef\tempurl%
\url{https://doi.org/10.1007/s11634-014-0192-4}
\showDOI{\tempurl}


\bibitem[Nakajima et~al\mbox{.}(2003)]%
        {NAM0}
\bibfield{author}{\bibinfo{person}{Y. Nakajima}, \bibinfo{person}{H. Kashioka}, \bibinfo{person}{K. Shikano}, {and} \bibinfo{person}{N. Campbell}.} \bibinfo{year}{2003}\natexlab{}.
\newblock \showarticletitle{Non-audible murmur recognition input interface using stethoscopic microphone attached to the skin}. In \bibinfo{booktitle}{\emph{2003 IEEE International Conference on Acoustics, Speech, and Signal Processing, 2003. Proceedings. (ICASSP '03).}}, Vol.~\bibinfo{volume}{5}. \bibinfo{pages}{V--708}.
\newblock
\urldef\tempurl%
\url{https://doi.org/10.1109/ICASSP.2003.1200069}
\showDOI{\tempurl}


\bibitem[Nam et~al\mbox{.}(2020)]%
        {unmasked}
\bibfield{author}{\bibinfo{person}{Hye~Yeon Nam}, \bibinfo{person}{Iyleah Hernandez}, {and} \bibinfo{person}{Brendan Harmon}.} \bibinfo{year}{2020}\natexlab{}.
\newblock \showarticletitle{Unmasked}. In \bibinfo{booktitle}{\emph{Adjunct Publication of the 33rd Annual ACM Symposium on User Interface Software and Technology}} (Virtual Event, USA) \emph{(\bibinfo{series}{UIST '20 Adjunct})}. \bibinfo{publisher}{Association for Computing Machinery}, \bibinfo{address}{New York, NY, USA}, \bibinfo{pages}{111–113}.
\newblock
\showISBNx{9781450375153}
\urldef\tempurl%
\url{https://doi.org/10.1145/3379350.3416137}
\showDOI{\tempurl}


\bibitem[Ozturk et~al\mbox{.}(2023)]%
        {RadioSES}
\bibfield{author}{\bibinfo{person}{Muhammed~Zahid Ozturk}, \bibinfo{person}{Chenshu Wu}, \bibinfo{person}{Beibei Wang}, \bibinfo{person}{Min Wu}, {and} \bibinfo{person}{K.~J.~Ray Liu}.} \bibinfo{year}{2023}\natexlab{}.
\newblock \showarticletitle{RadioSES: mmWave-Based Audioradio Speech Enhancement and Separation System}.
\newblock \bibinfo{journal}{\emph{IEEE/ACM Trans. Audio, Speech and Lang. Proc.}}  \bibinfo{volume}{31} (\bibinfo{date}{March} \bibinfo{year}{2023}), \bibinfo{pages}{1333–1347}.
\newblock
\showISSN{2329-9290}
\urldef\tempurl%
\url{https://doi.org/10.1109/TASLP.2023.3250846}
\showDOI{\tempurl}


\bibitem[Paliwal and Basu(1987)]%
        {kalmanfilter}
\bibfield{author}{\bibinfo{person}{K. Paliwal} {and} \bibinfo{person}{A. Basu}.} \bibinfo{year}{1987}\natexlab{}.
\newblock \showarticletitle{A speech enhancement method based on Kalman filtering}. In \bibinfo{booktitle}{\emph{ICASSP '87. IEEE International Conference on Acoustics, Speech, and Signal Processing}}, Vol.~\bibinfo{volume}{12}. \bibinfo{pages}{177--180}.
\newblock
\urldef\tempurl%
\url{https://doi.org/10.1109/ICASSP.1987.1169756}
\showDOI{\tempurl}


\bibitem[Radford et~al\mbox{.}(2023)]%
        {openai-whisper}
\bibfield{author}{\bibinfo{person}{Alec Radford}, \bibinfo{person}{Jong~Wook Kim}, \bibinfo{person}{Tao Xu}, \bibinfo{person}{Greg Brockman}, \bibinfo{person}{Christine McLeavey}, {and} \bibinfo{person}{Ilya Sutskever}.} \bibinfo{year}{2023}\natexlab{}.
\newblock \showarticletitle{Robust speech recognition via large-scale weak supervision}. In \bibinfo{booktitle}{\emph{Proceedings of the 40th International Conference on Machine Learning}} (Honolulu, Hawaii, USA) \emph{(\bibinfo{series}{ICML'23})}. \bibinfo{publisher}{JMLR.org}, Article \bibinfo{articleno}{1182}, \bibinfo{numpages}{27}~pages.
\newblock


\bibitem[Saeidi et~al\mbox{.}(2015)]%
        {Saeidi2015-dm}
\bibfield{author}{\bibinfo{person}{R. Saeidi}, \bibinfo{person}{T. Niemi}, \bibinfo{person}{H. Karppelin}, \bibinfo{person}{J. Pohjalainen}, \bibinfo{person}{T. Kinnunen}, {and} \bibinfo{person}{P. Alku}.} \bibinfo{year}{2015}\natexlab{}.
\newblock \showarticletitle{Speaker recognition for speech under face cover}. In \bibinfo{booktitle}{\emph{Proc. Interspeech 2015}}. \bibinfo{pages}{1012--1016}.
\newblock
\urldef\tempurl%
\url{https://doi.org/10.21437/Interspeech.2015-275}
\showDOI{\tempurl}


\bibitem[Sakashita et~al\mbox{.}(2016)]%
        {yadori}
\bibfield{author}{\bibinfo{person}{Mose Sakashita}, \bibinfo{person}{Keisuke Kawahara}, \bibinfo{person}{Amy Koike}, \bibinfo{person}{Kenta Suzuki}, \bibinfo{person}{Ippei Suzuki}, {and} \bibinfo{person}{Yoichi Ochiai}.} \bibinfo{year}{2016}\natexlab{}.
\newblock \showarticletitle{Yadori: Mask-Type User Interface for Manipulation of Puppets}. In \bibinfo{booktitle}{\emph{ACM SIGGRAPH 2016 Emerging Technologies}} (Anaheim, California) \emph{(\bibinfo{series}{SIGGRAPH '16})}. \bibinfo{publisher}{Association for Computing Machinery}, \bibinfo{address}{New York, NY, USA}, Article \bibinfo{articleno}{23}, \bibinfo{numpages}{1}~pages.
\newblock
\showISBNx{9781450343725}
\urldef\tempurl%
\url{https://doi.org/10.1145/2929464.2929478}
\showDOI{\tempurl}


\bibitem[Schilk et~al\mbox{.}(2023)]%
        {In-Ear-Voice}
\bibfield{author}{\bibinfo{person}{Philipp Schilk}, \bibinfo{person}{Niccol\`{o} Polvani}, \bibinfo{person}{Andrea Ronco}, \bibinfo{person}{Milos Cernak}, {and} \bibinfo{person}{Michele Magno}.} \bibinfo{year}{2023}\natexlab{}.
\newblock \showarticletitle{In-Ear-Voice: Towards Milli-Watt Audio Enhancement With Bone-Conduction Microphones for In-Ear Sensing Platforms}. In \bibinfo{booktitle}{\emph{Proceedings of the 8th ACM/IEEE Conference on Internet of Things Design and Implementation}} (San Antonio, TX, USA) \emph{(\bibinfo{series}{IoTDI '23})}. \bibinfo{publisher}{Association for Computing Machinery}, \bibinfo{address}{New York, NY, USA}, \bibinfo{pages}{1–12}.
\newblock
\showISBNx{9798400700378}
\urldef\tempurl%
\url{https://doi.org/10.1145/3576842.3582365}
\showDOI{\tempurl}


\bibitem[Schulte et~al\mbox{.}(2020)]%
        {Schulte2020-me}
\bibfield{author}{\bibinfo{person}{Antonia Schulte}, \bibinfo{person}{Rodrigo Suarez-Ibarrola}, \bibinfo{person}{Daniel Wegen}, \bibinfo{person}{Philippe-Fabian Pohlmann}, \bibinfo{person}{Elina Petersen}, {and} \bibinfo{person}{Arkadiusz Miernik}.} \bibinfo{year}{2020}\natexlab{}.
\newblock \showarticletitle{Automatic speech recognition in the operating room – An essential contemporary tool or a redundant gadget? A survey evaluation among physicians in form of a qualitative study}.
\newblock \bibinfo{journal}{\emph{Annals of Medicine and Surgery}}  \bibinfo{volume}{59} (\bibinfo{year}{2020}), \bibinfo{pages}{81--85}.
\newblock
\showISSN{2049-0801}
\urldef\tempurl%
\url{https://doi.org/10.1016/j.amsu.2020.09.015}
\showDOI{\tempurl}


\bibitem[Shimizu et~al\mbox{.}(2009)]%
        {nam_microphone}
\bibfield{author}{\bibinfo{person}{Shota Shimizu}, \bibinfo{person}{Makoto Otani}, {and} \bibinfo{person}{Tatsuya Hirahara}.} \bibinfo{year}{2009}\natexlab{}.
\newblock \showarticletitle{Frequency characteristics of several non-audible murmur (NAM) microphones}.
\newblock \bibinfo{journal}{\emph{Acoustical Science and Technology}} \bibinfo{volume}{30}, \bibinfo{number}{2} (\bibinfo{year}{2009}), \bibinfo{pages}{139--142}.
\newblock
\urldef\tempurl%
\url{https://doi.org/10.1250/ast.30.139}
\showDOI{\tempurl}


\bibitem[Subakan et~al\mbox{.}(2021)]%
        {sepformer}
\bibfield{author}{\bibinfo{person}{Cem Subakan}, \bibinfo{person}{Mirco Ravanelli}, \bibinfo{person}{Samuele Cornell}, \bibinfo{person}{Mirko Bronzi}, {and} \bibinfo{person}{Jianyuan Zhong}.} \bibinfo{year}{2021}\natexlab{}.
\newblock \showarticletitle{Attention Is All You Need In Speech Separation}. In \bibinfo{booktitle}{\emph{ICASSP 2021 - 2021 IEEE International Conference on Acoustics, Speech and Signal Processing (ICASSP)}}. \bibinfo{pages}{21--25}.
\newblock
\urldef\tempurl%
\url{https://doi.org/10.1109/ICASSP39728.2021.9413901}
\showDOI{\tempurl}


\bibitem[Sun and Zhang(2021)]%
        {Sun2021-dl}
\bibfield{author}{\bibinfo{person}{Ke Sun} {and} \bibinfo{person}{Xinyu Zhang}.} \bibinfo{year}{2021}\natexlab{}.
\newblock \showarticletitle{UltraSE: single-channel speech enhancement using ultrasound}. In \bibinfo{booktitle}{\emph{Proceedings of the 27th Annual International Conference on Mobile Computing and Networking}} (New Orleans, Louisiana) \emph{(\bibinfo{series}{MobiCom '21})}. \bibinfo{publisher}{Association for Computing Machinery}, \bibinfo{address}{New York, NY, USA}, \bibinfo{pages}{160–173}.
\newblock
\showISBNx{9781450383424}
\urldef\tempurl%
\url{https://doi.org/10.1145/3447993.3448626}
\showDOI{\tempurl}


\bibitem[Suzuki et~al\mbox{.}(2020)]%
        {mouthgesture}
\bibfield{author}{\bibinfo{person}{Yutaro Suzuki}, \bibinfo{person}{Kodai Sekimori}, \bibinfo{person}{Yuki Yamato}, \bibinfo{person}{Yusuke Yamasaki}, \bibinfo{person}{Buntarou Shizuki}, {and} \bibinfo{person}{Shin Takahashi}.} \bibinfo{year}{2020}\natexlab{}.
\newblock \showarticletitle{A Mouth Gesture Interface Featuring a Mutual-Capacitance Sensor Embedded in a Surgical Mask}. In \bibinfo{booktitle}{\emph{Human-Computer Interaction. Multimodal and Natural Interaction}}, \bibfield{editor}{\bibinfo{person}{Masaaki Kurosu}} (Ed.). \bibinfo{publisher}{Springer International Publishing}, \bibinfo{address}{Cham}, \bibinfo{pages}{154--165}.
\newblock
\showISBNx{978-3-030-49062-1}


\bibitem[Thibodeau et~al\mbox{.}(2021)]%
        {mask-is-difficult-to-speech-recognition1}
\bibfield{author}{\bibinfo{person}{Linda~M. Thibodeau}, \bibinfo{person}{Rachel~B. Thibodeau-Nielsen}, \bibinfo{person}{Chi Mai~Q. Tran}, {and} \bibinfo{person}{Ryan T.~S. Jacob}.} \bibinfo{year}{2021}\natexlab{}.
\newblock \showarticletitle{Communicating During COVID-19: The Effect of Transparent Masks for Speech Recognition in Noise}.
\newblock \bibinfo{journal}{\emph{Ear and Hearing}} \bibinfo{volume}{42}, \bibinfo{number}{4} (\bibinfo{date}{Jul} \bibinfo{year}{2021}), \bibinfo{pages}{772--781}.
\newblock
\urldef\tempurl%
\url{https://doi.org/10.1097/AUD.0000000000001065}
\showDOI{\tempurl}


\bibitem[Tipparaju et~al\mbox{.}(2020a)]%
        {Respiration}
\bibfield{author}{\bibinfo{person}{Vishal~Varun Tipparaju}, \bibinfo{person}{Di Wang}, \bibinfo{person}{Jingjing Yu}, \bibinfo{person}{Fang Chen}, \bibinfo{person}{Francis Tsow}, \bibinfo{person}{Erica Forzani}, \bibinfo{person}{Nongjian Tao}, {and} \bibinfo{person}{Xiaojun Xian}.} \bibinfo{year}{2020}\natexlab{a}.
\newblock \showarticletitle{Respiration pattern recognition by wearable mask device}.
\newblock \bibinfo{journal}{\emph{Biosensors and Bioelectronics}}  \bibinfo{volume}{169} (\bibinfo{year}{2020}), \bibinfo{pages}{112590}.
\newblock
\showISSN{0956-5663}
\urldef\tempurl%
\url{https://doi.org/10.1016/j.bios.2020.112590}
\showDOI{\tempurl}


\bibitem[Tipparaju et~al\mbox{.}(2020b)]%
        {ReliableBreathing}
\bibfield{author}{\bibinfo{person}{Vishal~Varun Tipparaju}, \bibinfo{person}{Xiaojun Xian}, \bibinfo{person}{Devon Bridgeman}, \bibinfo{person}{Di Wang}, \bibinfo{person}{Francis Tsow}, \bibinfo{person}{Erica Forzani}, {and} \bibinfo{person}{Nongjian Tao}.} \bibinfo{year}{2020}\natexlab{b}.
\newblock \showarticletitle{Reliable Breathing Tracking With Wearable Mask Device}.
\newblock \bibinfo{journal}{\emph{IEEE Sensors Journal}} \bibinfo{volume}{20}, \bibinfo{number}{10} (\bibinfo{year}{2020}), \bibinfo{pages}{5510--5518}.
\newblock
\urldef\tempurl%
\url{https://doi.org/10.1109/JSEN.2020.2969635}
\showDOI{\tempurl}


\bibitem[Toscano and Toscano(2021)]%
        {mask-is-difficult-to-speech-recognition2}
\bibfield{author}{\bibinfo{person}{Joseph~C. Toscano} {and} \bibinfo{person}{Caroline~M. Toscano}.} \bibinfo{year}{2021}\natexlab{}.
\newblock \showarticletitle{Effects of face masks on speech recognition in multi-talker babble noise}.
\newblock \bibinfo{journal}{\emph{PLOS ONE}} \bibinfo{volume}{16}, \bibinfo{number}{2} (\bibinfo{date}{Feb} \bibinfo{year}{2021}), \bibinfo{pages}{e0246842}.
\newblock
\urldef\tempurl%
\url{https://doi.org/10.1371/journal.pone.0246842}
\showDOI{\tempurl}


\bibitem[Veluri et~al\mbox{.}(2024)]%
        {Veluri2024-lp}
\bibfield{author}{\bibinfo{person}{Bandhav Veluri}, \bibinfo{person}{Malek Itani}, \bibinfo{person}{Tuochao Chen}, \bibinfo{person}{Takuya Yoshioka}, {and} \bibinfo{person}{Shyamnath Gollakota}.} \bibinfo{year}{2024}\natexlab{}.
\newblock \showarticletitle{Look Once to Hear: Target Speech Hearing with Noisy Examples}. In \bibinfo{booktitle}{\emph{Proceedings of the 2024 CHI Conference on Human Factors in Computing Systems}} (Honolulu, HI, USA) \emph{(\bibinfo{series}{CHI '24})}. \bibinfo{publisher}{Association for Computing Machinery}, \bibinfo{address}{New York, NY, USA}, Article \bibinfo{articleno}{37}, \bibinfo{numpages}{16}~pages.
\newblock
\showISBNx{9798400703300}
\urldef\tempurl%
\url{https://doi.org/10.1145/3613904.3642057}
\showDOI{\tempurl}


\bibitem[Vijayan et~al\mbox{.}(2017)]%
        {throatmic1}
\bibfield{author}{\bibinfo{person}{Amritha Vijayan}, \bibinfo{person}{Bipil~Mary Mathai}, \bibinfo{person}{Karthik Valsalan}, \bibinfo{person}{Riyanka~Raji Johnson}, \bibinfo{person}{Lani~Rachel Mathew}, {and} \bibinfo{person}{K. Gopakumar}.} \bibinfo{year}{2017}\natexlab{}.
\newblock \showarticletitle{Throat microphone speech recognition using mfcc}. In \bibinfo{booktitle}{\emph{2017 International Conference on Networks \& Advances in Computational Technologies (NetACT)}}. \bibinfo{pages}{392--395}.
\newblock
\urldef\tempurl%
\url{https://doi.org/10.1109/NETACT.2017.8076802}
\showDOI{\tempurl}


\bibitem[Wang et~al\mbox{.}(2022)]%
        {Wang2023-ll}
\bibfield{author}{\bibinfo{person}{Mou Wang}, \bibinfo{person}{Junqi Chen}, \bibinfo{person}{Xiao-Lei Zhang}, {and} \bibinfo{person}{Susanto Rahardja}.} \bibinfo{year}{2022}\natexlab{}.
\newblock \showarticletitle{End-to-End Multi-Modal Speech Recognition on an Air and Bone Conducted Speech Corpus}.
\newblock \bibinfo{journal}{\emph{IEEE/ACM Trans. Audio, Speech and Lang. Proc.}}  \bibinfo{volume}{31} (\bibinfo{date}{Nov.} \bibinfo{year}{2022}), \bibinfo{pages}{513–524}.
\newblock
\showISSN{2329-9290}
\urldef\tempurl%
\url{https://doi.org/10.1109/TASLP.2022.3224305}
\showDOI{\tempurl}


\bibitem[Wichern et~al\mbox{.}(2019)]%
        {Wichern2019WHAM}
\bibfield{author}{\bibinfo{person}{Gordon Wichern}, \bibinfo{person}{Joe Antognini}, \bibinfo{person}{Michael Flynn}, \bibinfo{person}{Licheng~Richard Zhu}, \bibinfo{person}{Emmett McQuinn}, \bibinfo{person}{Dwight Crow}, \bibinfo{person}{Ethan Manilow}, {and} \bibinfo{person}{Jonathan~Le Roux}.} \bibinfo{year}{2019}\natexlab{}.
\newblock \showarticletitle{{WHAM!: Extending Speech Separation to Noisy Environments}}. In \bibinfo{booktitle}{\emph{Proc. Interspeech}}. \bibinfo{pages}{1368--1372}.
\newblock
\urldef\tempurl%
\url{https://doi.org/10.21437/Interspeech.2019-2821}
\showDOI{\tempurl}


\bibitem[Yamamoto et~al\mbox{.}(2023)]%
        {masktrap}
\bibfield{author}{\bibinfo{person}{Takumi Yamamoto}, \bibinfo{person}{Katsutoshi Masai}, \bibinfo{person}{Anusha Withana}, {and} \bibinfo{person}{Yuta Sugiura}.} \bibinfo{year}{2023}\natexlab{}.
\newblock \showarticletitle{Masktrap: Designing and Identifying Gestures to Transform Mask Strap into an Input Interface}. In \bibinfo{booktitle}{\emph{Proceedings of the 28th International Conference on Intelligent User Interfaces}} (Sydney, NSW, Australia) \emph{(\bibinfo{series}{IUI '23})}. \bibinfo{publisher}{Association for Computing Machinery}, \bibinfo{address}{New York, NY, USA}, \bibinfo{pages}{762–775}.
\newblock
\showISBNx{9798400701061}
\urldef\tempurl%
\url{https://doi.org/10.1145/3581641.3584062}
\showDOI{\tempurl}


\bibitem[Yoshitaka et~al\mbox{.}(2005)]%
        {NAM1}
\bibfield{author}{\bibinfo{person}{NAKAJIMA Yoshitaka}, \bibinfo{person}{KASHIOKA Hideki}, \bibinfo{person}{CAMPBELL Nick}, {and} \bibinfo{person}{SHIKANO Kiyohiro}.} \bibinfo{year}{2005}\natexlab{}.
\newblock \showarticletitle{Non-Audible Murmur (NAM) Recognition}.
\newblock \bibinfo{journal}{\emph{IEICE TRANSACTIONS on Information and Systems}} \bibinfo{volume}{E89-D}, \bibinfo{number}{1} (\bibinfo{year}{2005}).
\newblock


\bibitem[Zelinka and Sigmund(2010)]%
        {Zelinka2010-rn}
\bibfield{author}{\bibinfo{person}{Petr Zelinka} {and} \bibinfo{person}{Milan Sigmund}.} \bibinfo{year}{2010}\natexlab{}.
\newblock \showarticletitle{Towards reliable speech recognition in operating room noise environment}. In \bibinfo{booktitle}{\emph{20th International Conference Radioelektronika 2010}}. \bibinfo{pages}{1--4}.
\newblock
\urldef\tempurl%
\url{https://doi.org/10.1109/RADIOELEK.2010.5478597}
\showDOI{\tempurl}


\bibitem[Zhang et~al\mbox{.}(2023)]%
        {Zhang2023-yu}
\bibfield{author}{\bibinfo{person}{Jun Zhang}, \bibinfo{person}{Jingyue Wu}, \bibinfo{person}{Yiyi Qiu}, \bibinfo{person}{Aiguo Song}, \bibinfo{person}{Weifeng Li}, \bibinfo{person}{Xin Li}, {and} \bibinfo{person}{Yecheng Liu}.} \bibinfo{year}{2023}\natexlab{}.
\newblock \showarticletitle{Intelligent speech technologies for transcription, disease diagnosis, and medical equipment interactive control in smart hospitals: A review}.
\newblock \bibinfo{journal}{\emph{Computers in Biology and Medicine}}  \bibinfo{volume}{153} (\bibinfo{year}{2023}), \bibinfo{pages}{106517}.
\newblock
\showISSN{0010-4825}
\urldef\tempurl%
\url{https://doi.org/10.1016/j.compbiomed.2022.106517}
\showDOI{\tempurl}


\bibitem[Zhang et~al\mbox{.}(2021)]%
        {Zhang2021-hs}
\bibfield{author}{\bibinfo{person}{Qian Zhang}, \bibinfo{person}{Dong Wang}, \bibinfo{person}{Run Zhao}, \bibinfo{person}{Yinggang Yu}, {and} \bibinfo{person}{Junjie Shen}.} \bibinfo{year}{2021}\natexlab{}.
\newblock \showarticletitle{Sensing to Hear: Speech Enhancement for Mobile Devices Using Acoustic Signals}.
\newblock \bibinfo{journal}{\emph{Proc. ACM Interact. Mob. Wearable Ubiquitous Technol.}} \bibinfo{volume}{5}, \bibinfo{number}{3}, Article \bibinfo{articleno}{137} (\bibinfo{date}{Sept.} \bibinfo{year}{2021}), \bibinfo{numpages}{30}~pages.
\newblock


\end{thebibliography}

\end{document}